
\input phyzzx.tex
\titlepage
\hfill{SISSA-ISAS 107/91/EP}
\title{Integrable Discrete Linear Systems
and One-Matrix Random Models}
\author{ L.Bonora}
\address{ International School for Advanced Studies (SISSA/ISAS),
Strada Costiera 11, I-34014 Trieste,
 Italy and INFN, sezione di Trieste}
\author{M. Martellini}
\address{Dipartimento di Fisica,
Universit\`a di Milano, I-20133 Milano, Italy and INFN,
sezione di Pavia, I-27100 Pavia, Italy}
\andauthor{ C.S.Xiong}
\address{ International School for Advanced Studies (SISSA/ISAS),
Strada Costiera 11, I-34014 Trieste, Italy}
\vskip .6truein
\abstract
In this paper we analyze one--matrix models by means of the associated
 discrete
linear systems. We see that
the consistency conditions of the discrete linear system lead to the
Virasoro constraints. The linear system is endowed with gauge

invariances.
We show that invariance under time--independent gauge transformations
entails the integrability of the model, while the double scaling limit
is connected with a time-dependent gauge transformation.
We derive the continuum version of the discrete linear system,
we prove  that  the partition function is actually
the $\tau$--function of the KdV hierarchy and that the linear system
completely determines the Virasoro constraints.

\vskip .4truein

\endpage
\pagenumber=1

\section{Introduction}

In the study of two dimensional quantum gravity,
one of the main approaches is provided by the random matrix models.
In this context the most interesting and calculable
model is the one Hermitian matrix model, which has been
studied intensively [1] (for reviews, see [2]). At the discrete level,
 it is
an integrable system [3], and its partition function
satisfies certain constraints, which belong to the Borel
subalgebra of the c=1 non--twisted Virasoro algebra[4]. After taking
the double scaling limit the system is expected to become
a KdV system restricted by twisted Virasoro constraints [5,6].
However, there is still a gap between
the discrete level and its continuum version:
first, the meaning of the Virasoro constraints is not clear;
secondly, the identification of the partition function with the
$\tau$--function  as well as the appearance
of the twisting has not been justified yet.

In this paper we will try to shed light upon these points
by analyzing an auxiliary tool in one-matrix models: the
relevant associated discrete linear systems.
We will show in fact, in section 2, that
a one--matrix model is equivalent to a certain discrete linear
system (DLS) in which the string equation appears as a compatibility
condition. In section 3 we will derive the discrete Virasoro conditions
from the consistency conditions of the DLS.
In section 4 we examine the ``time-independent" gauge symmetry
of the DLS and show that it is equivalent
to the integrability of the system. In section 5 another kind of
gauge symmetry is considered which includes in particular rescalings of
the couplings, and leads naturally to the double scaling limit.
In the last two sections we examine the continuum limit of the DLS
and study its properties. In section 6 we determine the continuum analog
of the DLS and find its consistency conditions. Finally, using the latter
exactly as in the discrete case, in section 7 we determine the continuum
Virasoro constraints. Our result is that
the partition function is actually described by a KdV--$\tau$ function,
and the constraints are the twisted ones. This result had only been
conjectured up to now.

\section{ One--Matrix Model}

In this section we review the main results concerning
the one--matrix model and introduce the associated discrete
linear system (DLS).

The
partition function of the one- matrix model is defined by [7]
$$Z_N(t)=\int\,dM e^{-TrV(M)}
=\int\prod_{i=1}^N d\lambda_i\Delta^2(\lambda)
\exp({-\sum_{i=1}^N V(\lambda_i)})
\eqno\hbox{(2.1)}$$
\noindent
with the potential
$$V(M)=\sum_{r=1}^{\infty}t_r M^r$$

A powerful tool to study this model is the use of orthogonal polynomials.
They are normalized as follows
$$P_n(\lambda)=\lambda^n+\ldots\eqno\hbox{(2.2)}$$
and satisfy the orthogonal and recursion relations
$$\int\,d\lambda e^{-V(\lambda)}P_n(\lambda)P_m(\lambda)
=\delta_{n,m}h_n(t)\eqno\hbox{(2.3)}$$
$$\lambda P_n(\lambda)=P_{n+1}(\lambda)+S_nP_n
(\lambda)+R_nP_{n-1}(\lambda)
\eqno\hbox{(2.4)}$$
\noindent
It is convenient to introduce another set of the polynomials
$$\xi_n(\lambda)\equiv{1\over{\sqrt{h_n}}}e^{-{1\over2}V(\lambda)}
P_n(\lambda)$$
\noindent
Then eqs. (2.3) and (2.4) become
$$\int\,d\lambda \xi_n(t,\lambda)\xi_m(t,\lambda)
=\delta_{n,m}\eqno\hbox{(2.5a)}$$
$$\lambda \xi_n=\sqrt{R_{n+1}} \xi_{n+1}
+S_n\xi_n
+\sqrt{R_{n}}\xi_{n-1}
\eqno\hbox{(2.5b)}$$
where the function $R_n(t)$ are defined as the ratio
$$R_n\equiv{{h_{n}}\over{h_{n-1}}},\quad\quad n\geq 1.$$
\noindent
One can show that
$$S_n=-{{\partial}\over{\partial t_1}}\ln{h_n(t)},\quad\quad n\geq 0
\eqno\hbox{(2.6a)}$$
\noindent
or more generally,
$${{\partial}\over{\partial t_r}}\ln{h_n(t)}=-(Q^r)_{nn}\qquad\qquad
n\geq 0\eqno\hbox{(2.6b)}$$
\noindent
where Q is the Jacobi matrix
$$\eqalign{Q_{nm}&\equiv
\int\,d\lambda \xi_m(t,\lambda)\lambda\xi_n(t,\lambda)
\cr\noalign{\vskip12pt}
&=\sqrt{R_{n+1}}\delta_{n,m-1}+S_n\delta_{n,m}
+\sqrt{R_n}\delta_{n,m+1}\cr}\eqno\hbox{(2.7)}$$
\noindent
In the same way, we introduce the matrix P
$$P_{nm}\equiv\int\,d\lambda \xi_m(\lambda)
{{\partial}\over{\partial {\lambda}}}
\xi_n(\lambda)\eqno\hbox{(2.8)}$$

Using the orthogonal polynomials,
 we can perform the integrations in (2.1)
and obtain
$$Z_N(t)=N!h_0h_1h_2\ldots h_{N-1}\qquad\qquad\quad\eqno\hbox{(2.9a)}$$
$${{\partial}\over{\partial t_r}}\ln{Z_N(t)}=-\sum_{n=0}^{N-1}Q^r_{nn}
\equiv-TrQ^r,\quad r\geq 1
\eqno\hbox{(2.9b)}$$
\noindent

Next we introduce the discrete linear system alluded to in the
introduction. Let us denote by  $\xi$ the column vector
with components $\xi_0, \xi_1, \xi_2, \ldots$ ,
use the recursion relations and differentiate the orthogonality
relations with respect to $t_r$: we arrive at the following
\underbar{discrete linear system} (DLS) of equations
$$Q\xi=\lambda\xi\eqno\hbox{(2.10a)}$$
$${{\partial}\over{\partial t_r}}\xi= Q^r_a\xi
\eqno\hbox{(2.10b)}$$
$${{\partial}\over{\partial{\lambda}}}\xi=P\xi
\eqno\hbox{(2.11a)}$$
$$P=\sum_{k=2}^{\infty}kt_kQ^{k-1}_a
\eqno\hbox{(2.11b)}$$
where the dependence on $t$ and $\lambda$ has been understood.
Here and throughout the paper we adopt the notation
$$(Q^r_a)_{nm}\equiv\cases{{1\over2}(Q^r)_{nm}, &$m< n$
\cr\noalign{\vskip12pt} 0,&m=n\cr\noalign{\vskip12pt}
-{1\over2}(Q^r)_{nm}, &$m>n$\cr}\eqno\hbox{(2.12)}$$
The product in the above equations is of course the matrix product.

The consistency conditions for this linear system give rise
to the discrete KdV hierarchy[8]
$${{\partial Q}\over{\partial t_r}}=[Q^r_a, Q]\eqno\hbox{(2.13)}$$
and to the so-called string equation
$$[Q, P]=1\eqno\hbox{(2.14)}$$
\noindent

All the integrability and criticality properties of the matrix model
are encoded in the DLS. Showing this is the aim of our paper.

It is interesting to answer the question:
to what extent is the correspondence
between discrete linear systems and one--matrix models one to one?
There certainly are linear systems that do not correspond to matrix models,
however if we impose the matrix $Q$ to have the Jacobi form (2.7), the
correspondence is one to one. Indeed let us start from the infinite column
vector $\xi$ with orthonormalized components $\xi_0,\xi_1,\xi_2,...$ as in
eq.(2.5a), and write the system
$$Q\xi=\lambda\xi\eqno\hbox{(2.10a')}$$
$${{\partial}\over{\partial t_r}}\xi= Q^r_a\xi
\eqno\hbox{(2.10b')}$$
Then we can reconstruct the partition function from
$${{\partial}\over{\partial t_r}}\ln{Z_N(t)}
=-TrQ^r,\quad r\geq 1$$
If we define now
$${{\partial}\over{\partial{\lambda}}}\xi=P\xi$$
we have the consistency condition
$${{\partial P}\over{\partial t_r}}= [Q^r_a, P]
\eqno\hbox{(2.11a)}$$
This admit the only solution
$$P=\sum_{k=2}^{\infty}kt_kQ^{k-1}_a
\eqno\hbox{(2.11b)}$$
using simply eqs.(2.10a') and (2.10b') beside the orthonormality conditions.

To end this section let us pause a bit to make a comment on
the DLS.
As we will see, it is justified to consider eq.(2.10a)
as a discrete version of the Schr\"odinger equation where
$\lambda$ plays the role of the spectral parameter;
while eq.(2.10b) shows the KdV--type of (isospectral)
deformations of the discrete Schr\"odinger equation we will be discussing
later on. It is well known that, typically, the Liouville theory
is characterized by a Schr\"odinger equation (actually by two of them,
one for each chirality). Even though we will be using throughout the paper
only the linear system (2.10-11), it is interesting to notice that we can push
this analogy further by introducing a discrete version of the Drinfeld-Sokolov
linear system. This is done as follows.

Introduce the two matrices $Q_1$ and $Q_2=(Q_1)^t$, $t$ meaning transposition,
by
$$
(Q_1)_{nm} = \sqrt{X_n} \delta_{n,m} + \sqrt{ {R_n}\over {{X_{n-1}}}}
\delta_{n,m+1} \eqno{(2.15)}
$$
and the block matrix and vector
$$
{\cal Q}=\left(\matrix {Q_1 & {-1}\cr
{-\lambda}& Q_2 \cr }\right),\quad\quad \quad\quad
\Xi = \left ( \matrix {\xi_1 \cr \xi_2\cr}\right) \eqno{(2.16)}
$$
Then the discrete Drinfeld-Sokolov linear system is
$$
{\cal Q}~\Xi =0 \eqno{(2.17)}
$$
This implies in particular
$
Q_2Q_1 \xi_1 = \lambda \xi_1,
$
and we recover eq.(2.10a), provided
$Q=Q_2Q_1$.
The latter condition allows us to uniquely
determine the $X_n$'s introduced above in terms of the $S_n$'s and $R_n$'s.
One finds
$$
X_0 = S_0, \quad\quad
X_n = {{Y_n}\over {Y_{n-1}}},\quad\quad n\geq 1 \eqno{(2.18)}
$$
where, introducing for the sake of homogeneity the notation $R_i=R_{ii-1}$,
$$
\eqalign{ Y_0 &= S_0 \cr
Y_1 &= S_1S_0 - R_{10} \cr
&\cdots\cr
Y_n &= \sum_{0\leq k\leq {n\over 2}}(-1)^k \sum_{i_1<i_2-1 <...<i_k-1}
S_n \cdots S_{i_k+1} R_{i_k i_k-1} S_{i_k-2}\cdots \cr
&~~~~~~~~~~~~~~~~~~\cdots S_{i_2+1} R_{i_2 i_2-1} S_{i_2-2}\cdots
S_{i_1+1} R_{i_1 i_1-1} S_{i_1-2}\cdots S_0 \cr}\eqno{(2.19)}
$$

We can go further. Let us introduce the $sl_2$ generators
$$
H=\left(\matrix {1& 0 \cr 0& -1\cr }\right), \quad\quad
E_+ = \left (\matrix {0&1 \cr 0&0 \cr }\right), \quad\quad
E_- = \left (\matrix {0&0 \cr 1&0 \cr }\right)
$$
while ${\bf 1}$ will denote the identity $2\times 2$ matrix. Next define
$$\eqalign{
\delta &= {{Q_1 +Q_2}\over 2} , \quad\quad \Delta = \delta {\bf 1}\cr
\pi &={{Q_1 -Q_2}\over 2} , \quad\quad \Pi = \pi H \cr
{\cal E}_+ &=E_++\lambda E_-, \quad\quad {\cal A}= \Pi -{\cal E}_+ \cr}
\eqno{(2.20)}
$$
Then the discrete Drinfeld-Sokolov linear system can be written
$$
(\Delta + {\cal A})~ \Xi=0 \eqno{(2.17')}
$$
We notice that ${\cal E}_+$ is the sum of the step operators corresponding
to the simple roots of the affine $sl_2$ algebra, provided we identify
the spectral parameter with the loop parameter. In this sense
${\cal A}$ can be thought of as the discrete analogue of an $sl_2$
loop algebra connection.

\vglue 0.5cm

\section{Virasoro constraints from DLS}

\vglue 0.5cm

An important piece of information for the matrix model is contained
in the so-called Virasoro constraints$^{[4,5,6]}$
$$L_nZ_N(t)=0,\qquad\quad\qquad\qquad\qquad\qquad n\geq -1\eqno\hbox{(3.1a)}$$
where
$$L_n=\sum_{k=1}^{\infty}kt_k{{\partial}\over{\partial t_{k-1}}}
-2N{{\partial}\over{\partial t_n}}+\sum_{k=1}^{n-1}
{{\partial}^2\over{\partial t_k\partial t_{n-k}}}
+N^2\delta_{n0,}\eqno\hbox{(3.1b)}$$
$$[L_n, L_m]=(n-m)L_{n+m}.\qquad\qquad\qquad\qquad\eqno\hbox{(3.1c)}$$
They completely determine the possible perturbations.

In this section we show that
the Virasoro constraints
result from the consistency conditions of the linear system (2.10-11),
eqs.(2.13-14).

To this end we rewrite the string
equation (2.14) in the following form
$$\sum_{k=2}^{\infty}kt_k{{\partial}\over{\partial t_{k-1}}}Q=-1,
\eqno\hbox{(3.2)}$$
\noindent
where we have used the KdV equations (2.13). Eq. (3.2) implies that
$$\ell S_n=-1,\qquad n\geq 0; \qquad\qquad
\ell\equiv \sum_{k=2}^{\infty}kt_k{{\partial}\over{\partial t_{k-1}}}
\eqno\hbox{(3.3)}$$
\noindent
which, by (2.6a), can be re--expressed as
$$\ell{{\partial}\over{\partial t_1}}\ln{h_n}=1, \quad\quad \forall n\geq 0$$
\noindent
since the operator $\ell$ commutes with ${{\partial}\over{\partial t_1}}$.
After
integrating over $t_1$, and using the formula (2.6b), we get
$$\sum _{k=1}^\infty kt_k (Q^{k-1})_{nn}+\alpha=0,\quad\quad \forall n\geq 0.
\eqno\hbox{(3.4)}$$
At first sight the integration constant $\alpha$ seems to depend on
$t_2, t_3,...$, but using the discrete KdV hierarchy and the string equation
one can prove that $\alpha$ is actually a constant.
After summation over n, from (2.9b) we obtain
$$(\ell -N(t_1-\alpha))Z_N(t)=0$$
We see that we can absorb $\alpha$ by a redefinition of $t_1$.
So finally we get
$$(\sum_{k=2}^{\infty}kt_k{{\partial}\over{\partial t_{k-1}}}-Nt_1)Z_N(t)=0
\eqno\hbox{(3.5)}$$
\noindent
which is nothing but the $L_{-1}$ constraint.

We remark that choosing even potentials would imply $S_n\equiv 0$; therefore
eq.(3.3) would be meaningless and would forbid us to recover the $L_{-1}$
Virasoro condition. We will see later on that in the continuum limit this
obstruction is removed.

In order to derive
the other Virasoro constraints, we introduce the quantities
$$B_n^{(r)}\equiv \sqrt{R_{n-1}\ldots R_{n-r}}P_{n,n-r},
\qquad r\geq 0.\eqno\hbox{(3.6)}$$
\noindent
Due to the string equation (2.14), one finds for the above symbols the
recursion
relations
$$\eqalign{B_{n+1}^{(r+1)}-B_n^{(r+1)}&
=(S_{n-r}-S_n)B_n^{(r)}+R_{n-r}B_n^{(r-1)}\cr\noalign{\vskip12pt}
&-R_{n-1}B_{n-1}^{(r-1)}
+\delta_{r,0}\cr}$$
\noindent
The first few ones are as follows
$$\eqalign{B_n^{(0)}&=0,\qquad\qquad B_n^{(1)}=n\cr\noalign{\vskip12pt}
B_n^{(2)}&=S_0+S_1+\ldots+S_{n-2}-(n-1)S_{n-1}\cr\noalign{\vskip12pt}
B_n^{(3)}&=\sum_{i=0}^{n-3}S^2_i+\sum_{i=0}^{n-4}S_iS_{i+1}
-\sum_{i=0}^{n-4}S_i(S_{n-2}+S_{n-1})\cr\noalign{\vskip12pt}
&-S_{n-3}S_{n-1}
+2\sum_{i=0}^{n-3}R_i-(n-2)R_{n-2}\cr\noalign{\vskip12pt}
&\qquad \qquad\ldots\ldots
\cr}\eqno\hbox{(3.7)}$$
\noindent
On the other hand, from the KdV equations, it is easy to see that
$${{\partial}\over{\partial t_1}}Tr Q =
{{\partial}\over{\partial t_1}}\sum_{i=0}^{N-1}S_i
=-R_{N-1}.\eqno\hbox{(3.8)}$$
\noindent
Furthermore, since $Q^r$ is a symmetric matrix, from the eqs.(2.11),
(3.4) with $\alpha=0$, one finds that
$$\sum_{k=1}^{\infty}kt_k
Q^{(k-1)}_{nl}=\cases{P_{nl}, &$n>l$\cr\noalign{\vskip12pt}
0,&n=l\cr\noalign{\vskip12pt}
-P_{nl}, &$n<l$\cr}.\eqno\hbox{(3.9)}$$
\noindent
Therefore, for any positive integer r, one obtains
$$\eqalign{\sum_{k=1}^{\infty}kt_k Q^{(k+r)}_{nn}&=
\sum_{k=1}^{\infty}\sum_{l=n-r-1}^{n+r+1}(kt_kQ^{(k-1)})_{nl}
(Q^{r+1})_{ln}\cr\noalign{\vskip12pt}
&=B_{n+r+1}^{(r+1)}+B_n^{(r+1)}+\ldots\cr}.\eqno\hbox{(3.10)}$$
\noindent
Then, after summation over n, the above equations give
the $L_r$--constraint.
For instance, for the first three cases, it is easy to check that
$$\eqalign{\sum_{k=1}^{\infty}kt_kQ^k_{nn}&=
\sum_{k=1}^{\infty}kt_kQ^{k-1}_{n,n-1}
Q_{n-1,n}+\sum_{k=1}^{\infty}kt_kQ^{k-1}_{nn}Q_{nn}+
\sum_{k=1}^{\infty}kt_kQ^{k-1}_{n,n+1}Q_{n+1,n}\cr\noalign{\vskip12pt}
&=B^{(1)}_n+B^{(1)}_{n+1}=2n+1\cr\noalign{\vskip12pt}
\sum_{k=1}^{\infty}kt_kQ^{k+1}_{nn}
&=B^{(1)}_{n+1}(S_n+S_{n+1})+B^{(1)}_n(S_n+S_{n-1})+B^{(2)}_n+
B^{(2)}_{n+2}\cr\noalign{\vskip12pt}
&=2[S_0+S_1+\ldots+S_{n-1}+(n+1)S_n]\cr\noalign{\vskip12pt}
\sum_{k=1}^{\infty}kt_kQ^{k+2}_{nn}
&=B^{(3)}_{n+3}+B^{(3)}_n+B^{(2)}_n(S_n+S_{n-1}+S_{n-2})+
B^{(2)}_{n+2}(S_n+S_{n+1}+S_{n+2})
\cr\noalign{\vskip12pt}
&+B^{(1)}_n(R_n+R_{n-1}+R_{n-2}+S^2_{n-1}+S_nS_{n-1}+S^2_n)
\cr\noalign{\vskip12pt}
&+B^{(1)}_{n+1}(R_n+R_{n-1}+R_{n+1}+S^2_{n+1}+S_nS_{n+1}+S^2_n)\cr},
$$
\smallskip
\noindent
which after summing over n, and making use of (2.9b), (3.7) and (3.8),
become
the Virasoro constraints
$$L_nZ_N(t)=0, \qquad\qquad n=0, 1, 2\eqno\hbox{(3.11a)}$$
$$L_n=\sum_{k=1}^{\infty}kt_k{{\partial}\over{\partial t_{k+n}}}
-2N{{\partial}\over{\partial t_n}}+\sum_{k=1}^{n-1}
{{\partial}^2\over{\partial t_k\partial t_{n-k}}}+N^2\delta_{n,0}.
\eqno\hbox{(3.11b)}$$
\noindent
The Virasoro algebraic structure (3.1c) ensures that the
higher order constraints are also true.

\section{Gauge Symmetry and Integrability}
\vglue 0.5cm

As we anticipated in the introduction
the discrete linear system (2.10) is characterized by gauge symmetries
which, on the one hand, ensure integrability and, on the other hand, allow us
to envisage the double scaling limit as a singular gauge transformation.
This section and the following one are devoted to studying the properties
of these gauge tranformations.

Let us consider the following transformation (at fixed $t_k$'s)
$$\cases{\xi\longrightarrow {\hat\xi}=G^{-1}\xi,\cr\noalign{\vskip12pt}
Q\longrightarrow {\hat Q}=G^{-1}QG\cr}\eqno\hbox{(4.1)}$$
\noindent
where G is a unitary matrix. If the transformed Jacobi matrix ${\hat Q}$
has the same structure as Q, i.e. only the diagonal line
and the first two off--diagonal lines are non--zero, and, moreover, if
$$\cases{{\hat Q}{\hat \xi}=\lambda{\hat \xi}\cr\noalign{\vskip12pt}
{{\partial}\over{\partial t_r}}{\hat \xi}={\hat Q}^r_a{\hat \xi}\cr}
\eqno\hbox{(4.2)}$$
\noindent
then, we say that our linear system is gauge invariant.

Let us examine these transformations more closely by considering
the infinitesimal transformation
$$G=1+\varepsilon g.$$
\noindent
Then, the invariance requires the matrix g to satisfy the
equations
$${\hat Q}=Q+\varepsilon [Q,g]\eqno\hbox{(4.3a)}$$
$${{\partial}\over{\partial t_r}}g=[Q^r_a, g]-[Q^r, g]_a.\eqno\hbox{(4.3b)}$$
\noindent
A non--trivial solution is
$$g=\sum_{k}b_kQ^k_a,\eqno\hbox{(4.4)}$$
\noindent
where $b_k$'s are time--independent constants. By abuse of language
we will call this a ``time--independent
gauge transformation''.

Let us consider the case when only
$b_k$ is nonzero. Then
$$\delta Q={\hat Q}-Q=\varepsilon b_k [Q, Q^k_a]=-\varepsilon b_k
{{\partial}\over{\partial t_k}}Q.\eqno\hbox{(4.5)}$$
\noindent
This corresponds to the transformation $t_k \rightarrow  t_k -\varepsilon b_k
$,
which
can be rephrased by saying that
the tuning of the time parameters is realized by
means of the gauge transformation (4.4).

This transformation has remarkable properties. On the one hand it can be
considered as
the discrete version of the conformal transformations, on the other hand
it leads to
the integrability of the linear system (and consequently to that of the
one--matrix model).

Let us consider in detail the latter claim. We can think of
$\delta Q$ given by eq.(4.5) as originating from
a Poisson bracket in the following
sense:
$$\delta Q\equiv \varepsilon \{ A_r, Q\}
\eqno\hbox{(4.6)}$$
\noindent
That is
$$ \{ A_r, Q\}\equiv[Q^r_a, Q].\eqno\hbox{(4.7)}$$
\noindent
where $A_r$ represents a Hamiltonian to be determined.
For the Hamiltonians we make the ansatz
$$
H_r\equiv {1\over r}\sum _{n=0}^\infty Q^r_{nn}\qquad\qquad
r=1,2,\ldots \eqno\hbox{(4.8)}
$$
and corresponding to the choice
$A_r=H_r, H_{r-1}, H_{r-2},...$, we obtain different Poisson brackets.
More explicitly we write
$$
\{H_{r-k+1} , Q \}_k = [Q_0^r,Q] , \eqno (4.9)
$$
In the following we will study in detail only the cases $k=1,2,3$.
Comparing the LHS with the RHS which can be explicitly calculated, we can
obtain the Poisson brackets for $R_i$ and $S_i$. Of course we have no
a priori guarantee that the brackets so obtained satisfy the Jacobi identity.
This has to be checked a posteriori.

While explicitly working out  the Poisson brackets one realizes that there
are two distint regimes according to whether $S_i =0$ or $\neq 0$.

i) {\it First regime}, i.e. $S_i=0$. We find two meaningful Poisson
brackets:
$$
\{ R_i, R_j\}_1 =R_iR_j (\delta_{j,i+1}-\delta_{i,j+1}) \eqno (4.10)
$$
and
$$
\eqalignno{\{ R_i, R_j\}_3= & R_iR_j(R_i+R_j)(\delta_{j,i+1}-\delta_{i,j+1})
\qquad\qquad\cr\noalign{\vskip12pt}
&+R_jR_{j-1}R_{j-2}\delta_{j,i+2}-R_iR_{i-1}R_{i-2}\delta_{i,j+2}
&(4.11)\cr}
$$
while
$$
\{R_i,R_j\}_2 \equiv 0
$$

ii) {\it Second regime}, i.e. $S_i\neq 0$.  We have two Poisson brackets:
$$
\eqalignno{\{ R_i, R_j\}_1 =&R_iR_j (\delta_{j,i+1}-\delta_{i,j+1})
& (4.12a)\cr\noalign{\vskip12pt}
\{ R_i, S_j\}_1=&R_iS_j(\delta_{j,i+1}-\delta_{i,j}) & (4.12b)
\cr\noalign{\vskip12pt}
\{ S_i, S_j\}_1=&R_i\delta_{j,i+1}-R_j\delta_{i,j+1}.
& (4,12c)\cr}
$$
and
$$
\eqalignno{\{R_i,R_j\}_2 =& 2 R_iR_j ( S_i \delta _{i,j-1}- S_j \delta_{i,j+1})
&(4.13a)\cr
\{R_i , S_j\}_2 =& R_i R_j(\delta _{i,j-1} + \delta _{i,j}) -R_iR_{j+1}
( \delta _{i,j+1}+ \delta _{i,j+2}) \cr
&+ R_i S_j^2(\delta_{i,j}- \delta_{i,j+1}) &(4.13b)\cr
\{S_i,S_j\}_2 =& (S_i+S_j) (R_j \delta_{i,j-1}- R_i \delta_{i,j+1})&(4.13c)\cr}
$$
For $k=3$ eq.(4.9) does not define a consistent bracket.

Let us conclude this section with a few remarks.
Due to eq.(4.7) one would expect all the Hamiltonians to
commute
$$\{H_n, H_m\}=0.\eqno\hbox{(4.14)}$$
for any meaningful Poisson structure.
However due to the subtleties connected with traces of infinite matrices, one
should verify this property starting from the Poisson brackets (4.9) and
(4.10).
We have done it for the first few cases.

Moreover, from the definition of the Poisson brackets, we know that
$$\{H_{r+2}, Q\}_1=\{H_r, Q\}_3$$
in the first regime
and
$$\{H_{r+1}, Q\}_1=\{H_r, Q\}_2$$
in the second, namely
the two Poisson structures are compatible with each other.
That is to say, in both regimes
the linear system has infinitely many conserved quantities and possesses a
bi--Hamiltonian structure.

Finally the Poisson brackets (4.10) and (4.11) are the same we come across
in the lattice version of the Liouville model [9]. In that case $R_i$ plays
the role of a lattice deformation of the classical Virasoro generators.
However one should remember that
in the first regime we cannot impose the string equation (see remark after
eq.(3.5)). So we find a remarkable similarity of structures
in the Liouville model on the lattice and in the one-matrix model
unconstrained by the string equation.
Strictly speaking, the Poisson structures of the one-matrix model are those
of the second regime. It is an interesting open problem what field theory on
the lattice they correspond to.

\noindent
\section{Reparametrization and Time-Dependent Gauge Transformations}
\vglue 0.5cm

The DLS (2.10) is form invariant under reparametrization of the $t_k$
couplings. That is
$$Q(\tilde t)\xi(\tilde t)=\lambda\xi(\tilde t),\quad\quad\quad
{{\partial}\over{\partial {\tilde t}_r}}\xi(\tilde t)=
(Q^r_a)(\tilde t)\xi(\tilde t)
$$
where $\tilde t =(\tilde t_1, \tilde t_2, ...)$ and $\tilde t_k$ is a
smooth functions of the $t_k$'s.

This invariance is a formal one. However, by combining gauge and
reparametrization transformations, we can obtain significant symmetries of
the system. Let us consider the transformations
$$
\cases{\cases{\tilde t_k = t_k+ \varepsilon (k-n) t_{k-n},
\quad\quad \forall k >n,
\quad\quad n\geq -1\cr
\tilde t_n = t_n - 2 N \varepsilon, \quad\quad n \geq 1\cr}\cr
\hat \xi(\tilde t)=G^{-1}(\tilde t)\xi (\tilde t),\cr
{\hat Q}(\tilde t)=G^{-1}(\tilde t)Q(\tilde t)G(\tilde t)\cr}\eqno\hbox{(5.1)}
$$
In this kind of setup it is possible to find $G= 1+\varepsilon g$ so that
$$
\hat Q(\tilde t) = Q(t)  \eqno\hbox{(5.2)}
$$
and the linear system becomes
$$
\cases{{Q}(t){\hat \xi}(\tilde t)=
\lambda{\hat \xi}(\tilde t)\cr\noalign{\vskip12pt}
{{\partial}\over{\partial {t}_r}}{\hat \xi}(\tilde t)=
{Q}^r_a(t){\hat \xi}(\tilde t)\cr}
\eqno\hbox{(5.3)}
$$
We refer to these as time-dependent gauge transformations.

Let us consider two examples
$$
\eqalign{
g&= P, \quad\quad \quad n=-1 \cr
g&= QP, \quad\quad\quad n=0 \cr}\eqno\hbox{(5.4)}
$$
In these two cases eq.(5.2) is satisfied (the linear system is invariant
but remark that $\xi$ is ${\underline {\rm not}}$ invariant for $n=0$)
and this fact leads straightforwardly
to the $L_{-1}$ and $L_0$ Virasoro constraints. We could as well obtain the
other Virasoro constraints, but in these cases the matrix $g$ has
a complicated form and will not be written down here.

\section{The Double Scaling Limit}

The purpose of this section is to recover a continuum version of the
DLS (2.10) in the double scaling limit. In general we
will exploit the idea that
the double scaling limit is mimicked by a finite version of
the transformation (5.1) above when $n=0$:
$$
t_r\longrightarrow \gamma^{r} t_r, \quad\quad \forall r \eqno\hbox{(6.1)}
$$
where $\gamma$ is a finite constant.
We recall that under this transformation $Q$ remains invariant.
In other words, the double scaling limit
is connected with a singular case of a symmetry operation on our DLS.

Let us consider a k-th order critical point and define, as usual,
the continuum variables
$$x\equiv {n\over{\beta}},\qquad R(x)\equiv R_n,\qquad
\xi(x)\equiv \xi_n.$$
\noindent
Moreover we set
$$\epsilon\equiv ({1\over{\beta}})^{{1\over{2k+1}}}, \qquad
{\tilde t_0}=(1-{n\over{\beta}})\beta^{{{2k}\over{2k+1}}}, \qquad
\partial\equiv {{\partial}\over{\partial {\tilde t_0}}},
\eqno\hbox{(6.2)}$$
\noindent
The double scaling limit corresponds to
$$
\beta \rightarrow \infty, \quad\quad N \rightarrow\infty,
\quad\quad \tilde t_0~~{\rm fixed}
$$
For large $n\sim N$ one has
$$x=1-\epsilon^{2k}{\tilde t_0},\qquad\quad
R(x)=1+\epsilon^2u({\tilde t_0})\eqno\hbox{(6.3)}$$
\noindent
where $u({\tilde t_0})$ is the specific heat.

The latter ansatz requires a comment. Let us consider the string equation
(2.14), suitably rescaled
$$
[Q, \bar P]={1\over \beta} ,\quad\quad\quad
\bar P= \sum_{r=2}^\infty r \bar t_r Q_a^{r-1}, \quad\quad\quad t_r = \beta
\bar t_r \eqno\hbox{(6.4)}
$$
where $\bar t_r$ are renormalized coupling constants (see below).
This equation establishes strong restrictions between the limiting
expressions of $Q$ and $\bar P$. With the simplifying assumption $t_{2r+1}=0$
{}~~$\forall r$, the above ansatz is correct, as is well known;
if we switch on the odd interactions the analysis is more complicated, the
above ansatz is still correct but we have not been able to
exclude other solutions. In the following we will stick to the case
of even potentials and to (6.3).

Let us now write down a few expansions which will be useful in the following.
It is easy to see that
$$R_{n\mp 1}=R(x\mp {1\over{\beta}})=1+\epsilon^2u({\tilde t_0}\pm \epsilon)
$$
$$\xi_{n\mp 1}=\xi(x\mp {1\over{\beta}})=
e^{\pm\epsilon\partial}\xi({\tilde t_0}).
$$
\noindent
Then, using Taylor expansion, we have the following formulas
$$R_{n+b}=1+\epsilon^2u-b\epsilon^3u^{'}+{{b^2}\over2}\epsilon^4u^{''}
-{{b^3}\over6}\epsilon^5u^{'''}+\ldots\eqno\hbox{(6.5a)}$$
$$\eqalign{\prod_{i=0}^q R_{n+b+i}&
=1+(q+1)\epsilon^2u-{1\over2}(q+1)(2b+q)\epsilon^3u^{'}\cr\noalign{\vskip12pt}
&+\epsilon^4\bigg(\sum_{i=0}^q{(b+i)^2\over2}u^{''}+{{(q+1)q}\over2}u^2\bigg)
\cr\noalign{\vskip12pt}
&-\epsilon^5\bigg(\sum_{i=0}^q{(b+i)^3\over6}u^{'''}
+{{(2b+q)(q+1)q}\over2}uu^{'}\bigg)+\ldots\cr}\eqno\hbox{(6.5b)}$$
$$\eqalign{\prod_{i=0}^q\sqrt{ R_{n+b+i}}&=1+{{q+1}\over2}\epsilon^2u
-{{(q+1)(2b+q)}\over4}\epsilon^3u^{'}\cr\noalign{\vskip12pt}
&+\epsilon^4\bigg(\sum_{i=0}^q{(b+i)^2\over4}u^{''}
+{{(q+1)(q-1)}\over8}u^2\bigg)\cr\noalign{\vskip12pt}
&-\epsilon^5\bigg(\sum_{i=0}^q{(b+i)^3\over12}u^{'''}
+{{(2b+q)(q+1)(q-1)}\over8}uu^{'}\bigg)+\ldots\cr}\eqno\hbox{(6.5c)}$$

Using these formulas one can derive
$$\eqalignno{Q&=
2+\epsilon^2(\partial^2+u)+{\cal O}(\epsilon^3), &(6.6a)\cr\noalign{\vskip12pt}
-Q_a&=
\epsilon\partial+{1\over6}\epsilon^3(\partial^3+3u\partial+{3\over2}u^{'})
+{1\over8}\epsilon^4(2u^{'}\partial+u^{''})\cr\noalign{\vskip12pt}
&+{1\over8}\epsilon^5\bigg({1\over15}\partial^5+{2\over3}u\partial^3+u^{'}
\partial^2+u^{''}\partial-u^2\partial-uu^{'}
+{1\over3}u^{'''}\bigg)+\ldots,&(6.6b)\cr\noalign{\vskip12pt}
-Q^2_a
&=2\epsilon\partial
+{4\over3}\epsilon^3(\partial^3+{3\over2}u\partial+{3\over4}u^{'})
+{1\over2}\epsilon^4(2u^{'}\partial+u^{''})\cr\noalign{\vskip12pt}
&+\epsilon^5\bigg({4\over15}\partial^5+{4\over3}u\partial^3+2u^{'}
\partial^2+{3\over2}u^{''}\partial
+{5\over12}u^{'''}\bigg)+\ldots,&(6.6c) \cr\noalign{\vskip12pt}
-Q^3_a
&=6\epsilon\partial+\epsilon^3(5\partial^3+9u\partial+{9\over2}u^{'})
+{9\over 2}\epsilon^4(2u^{'}\partial+u^{''})
+{{\epsilon^5}\over 2}\bigg({41\over10}\partial^5\cr\noalign{\vskip12pt}
&+15u\partial^3+{45\over2}u^{'}
\partial^2+{37\over2}u^{''}\partial+{9\over2}u^2\partial+{9\over2}uu^{'}
+{11\over2}u^{'''}\bigg)+\ldots,&(6.6d)\cr\noalign{\vskip12pt}
-Q^4_a
&=12\epsilon\partial+16\epsilon^3(\partial^3+{3\over2}u\partial+{3\over4}u^{'})
+6\epsilon^4(2u^{'}\partial+u^{''})
+2\epsilon^5\bigg({24\over5}\partial^5\cr\noalign{\vskip12pt}
&+16u\partial^3+24u^{'}
\partial^2+19u^{''}\partial+6u^2\partial+6uu^{'}
+{11\over2}u^{'''}\bigg)+\ldots,&(6.6e)\cr\noalign{\vskip12pt}
-Q^5_a
&=30\epsilon\partial+\epsilon^3(45\partial^3+75u\partial+{75\over2}u^{'})
+75\epsilon^4(2u^{'}\partial+u^{''})
+{5\over4}\epsilon^5(29\partial^5\cr\noalign{\vskip12pt}
&+90u\partial^3+135u^{'}
\partial^2+111u^{''}\partial
+45u^2\partial+45uu^{'}
+33u^{'''})+\ldots&(6.6f)\cr}
$$
etc.

Let us see now some consequences of the above expansions.
First of all let us notice that in the continuum limit the reduction
to even potentials does not contradict the string equation
as in the
discrete case. We should remember that the contradiction is exposed in
eq.(3.3). From the above expansions it is not difficult to see that in the
continuum limit it does not make sense to single out an equation like (3.3),
while the LHS of the string equation
is replaced by a differential operator even if
$t_{2r+1}=0~~\forall r$. So in the continuum limit on can safely choose
an even potential.

Next we consider the continuum limit of (2.10a). From eq.(6.6a) we see that
in a neighbourhood of the critical point $\lambda \sim 2$. Therefore
we introduce the renormalized quantities
$$
{\tilde \lambda}\equiv \epsilon^{-2}(\lambda-2),\qquad \qquad
{\tilde Q}\equiv  \partial^2+u \eqno\hbox{(6.7a)}
$$
$$
{{\partial}\over{\partial{\tilde \lambda}}}\xi={\tilde P}\xi,\qquad
\qquad {\tilde P}=\epsilon^2P=\epsilon^{1-2k}\bar P.\eqno\hbox{(6.7b)}
$$
Then, the discrete  Schr\"odinger equation goes over to its continuum version
$$
(\partial^2+u){\tilde\xi({\tilde t_0})}
={\tilde \lambda}{\tilde\xi({\tilde t_0})}\eqno\hbox{(6.8)}
$$
Similarly the string equation becomes
$$
[{\tilde Q}, {\tilde P}]=1.\eqno\hbox{(6.9)}
$$

We are now in a condition to explicitly determine critical points.
{}From eq.(2.11b) and (6.7b) (recall that we are working
with the simplifying assumption of even potential) we have
$$\eqalign{
{\tilde P}
&=\epsilon^{2}(2t_2Q_a+4t_4Q^3_a+6t_6Q^5_a+\ldots)\cr
&=\epsilon^{1-2k}(2\bar t_2Q_a+4\bar t_4Q^3_a+6\bar t_6Q^5_a+\ldots)\cr}
\eqno\hbox{(6.10)}
$$
We remarked above that the string equation (6.9) puts severe restrictions
not only in the discrete case but also in the double scaling limit.
{}From eq.(6.6) and (6.7), we see in particular that
all the operators $Q^r_a$'s are vanishing
in the limit $\epsilon\rightarrow$ 0, so that if one wants the string equation
to be satisfied, one must
let a certain subset of bare coupling constants in $P$ go to infinity (DSL).
The practical recipe is to look for combinations of $\bar t_{2r}$'s such that
all the singular terms in the second expression of (6.10) vanish.
Let us see a few significant examples.

i) k=2.
In this case only $\bar t_2$ and  $\bar t_4$ are nonzero, and $\beta
=\epsilon^{-5}$.
Then, from (6.6b,d) we see that only if we set
$$
\cases{t_2={15\over8}\epsilon^{-5}{\tilde t_2}
=\epsilon^{-5}{\bar t_2},\cr\noalign{\vskip12pt}
t_4=-{5\over32}\epsilon^{-5}{\tilde t_2}=\epsilon^{-5}{\bar t_4},\cr}
$$
are we able to eliminate the $\epsilon^{-1}\partial$ term in $\tilde P$, and
we get the known operator
$$
{\tilde P}={5\over2}
{\tilde t_2}(\partial^2+u)^{3\over2}_++{\cal O}(\epsilon)
$$
The string equation becomes
$$
-{5\over2}{\tilde t_2}{\cal R}^{'}_2[u]=1
$$
where we have introduced the Gelfand--Dickii polynomials
$$
{\cal R}^{'}_k[u]\equiv
[(\partial^2+u)^{k-{1\over2}}_+, \partial^2+u].\eqno\hbox{(6.11)}
$$
As is well-known at the critical point ${\tilde t_2}^c={8\over15}$,
the above string equation is the Painlev\'e equation of first kind
$${\tilde t_0}={1\over3}u^{''}+u^2.$$

ii) k=3. Only $t_2,t_4$ and $t_6$ are non-vanishing, $\beta =\epsilon ^{-7}$.
According to the above recipe we kill all the negative powers of $\epsilon$
in $\tilde P$ if we put
$$
\cases{t_2=-{105\over32}\epsilon^{-7}{\tilde t_3}=\epsilon^{-7}{\bar t_2},\cr
t_4={35\over64}\epsilon^{-7}{\tilde t_3}=\epsilon^{-7}{\bar t_4},\cr
t_6=-{7\over192}\epsilon^{-7}{\tilde t_3}=\epsilon^{-7}{\bar t_6},\cr}
$$
It is straightforward to show that
$${\tilde P}={7\over2}
{\tilde t_3}(\partial^2+u)^{5\over2}_++{\cal O}(\epsilon)
$$
and the string equation becomes
$$
-{7\over2}{\tilde t_3}{\cal R}^{'}_3[u]=1.$$
\noindent
The third critical point is at
${\tilde t_3}^c=-{16\over35}$ and
 the differential equation is
$${\tilde t}_0=-(u^3-{1\over2}{u^{'}}^2-uu^{''}+{1\over10}u^{(4)}).$$

One can proceed in this way and determine higher order critical points.
As is well-known, on a very general ground the form of the operator
${\tilde P}$ must be as follows
$${\tilde P}=\sum_{n=1}^{\infty}(n+{1\over2}){\tilde t_n}{\tilde Q}^{n+
{1\over2}}_++{\cal O}(\epsilon).\eqno\hbox{(6.12)}
$$
We conjecture that this form is induced by the following
coupling redefinitions:
$$t_{2r}=-\sum_{n=r}^{\infty}(n+{1\over2}){\tilde t_n}C_n a^{(n)}_r
\epsilon^{-(2n+1)}\equiv\sum_{n=r}^{\infty}\Gamma^n_r{\tilde t_n}
\eqno\hbox{(6.13)}$$
where
$$a^{(n)}_r=(-1)^{r+1}{{n!(r-1)!}\over{(n-r)!(2r)!}}$$
We checked eq.(6.13) for the first few cases and found
$$
C_1={1},\quad C_2=-{3\over4},\quad C_3={5\over8},\quad C_4=-{{35}\over {64}},
\ldots
$$
In general one has
$$C_n=(-1)^{n+1}{{(2n-1)!!}\over{2^{n-1}n!}}.
$$
These factors are determined in such a way as to reproduce the standard
KdV hierarchy.

It is worth noticing that 1) in eq.(6.12) and (6.13) we are considering all
the critical points at a time, and 2) the time transformation (6.13) is
made of a reparametrization plus a scale transformation of the type (6.1).

What is left for us to do is to analyze the continuum limit of the KdV
hierarchy. On the basis of eq.(6.13) one naively has
$$
{{\partial}\over{\partial{\tilde t_n}}}=-\sum_{r=1}^n(n+{1\over2})
C_n\epsilon^{-(2n+1)}a^{(n)}_r
{{\partial}\over{\partial{t_{2r}}}}.\eqno\hbox{(6.14)}
$$
So, in particular, on the basis of (2.10) and (6.6) we must have
$$
{{\partial}\over{\partial{\tilde t_1}}}\xi=
\bigg({3\over2}\epsilon^{-2}\partial+(\partial^2+u)^{3\over2}_+
+{\cal O}(\epsilon)\bigg)\xi$$
$${{\partial}\over{\partial{\tilde t_2}}}\xi=\bigg(-{15\over8}\epsilon^{-4}
\partial+(\partial^2+u)^{5\over2}_++{\cal O}(\epsilon)\bigg)\xi$$
$${{\partial}\over{\partial{\tilde t_3}}}\xi=\bigg({35\over32}
\epsilon^{-6}\partial+(\partial^2+u)^{7\over2}_+
+{\cal O}(\epsilon)\bigg)\xi$$
etc. These are however naive formulae since $\{\tilde t_n,~~n\geq 1\}$ is not
a complete set of parameters after we take the continuum limit. To correct
this we have to allow also for a $\partial$--dependent term in the RHS of
(6.14). This additional term takes care exactly of the divergent terms
(in the $\epsilon \rightarrow 0$ limit) in the RHS of the above equations.

Finally the evolution equations become
$${{\partial}\over{\partial{\tilde t_n}}}{\tilde\xi}
=(\partial^2+u)^{n+{1\over2}}_+{\tilde\xi},\quad\quad n\geq 0
\eqno\hbox{(6.15)}$$
\noindent
which result in the standard KdV flow
$${{\partial}\over{\partial{\tilde t_n}}}u=
[(\partial^2+u)^{n+{1\over2}}_+, \partial^2+u],\quad\quad n\geq 0
\eqno\hbox{(6.16)}$$

In eq.(6.15) $\tilde \xi$ is the limit of $\xi$ possibly multiplied by an
$\epsilon$--dependent factor which may be necessary in order to obtain a finite
result.

\section{The Virasoro Constraints in the Continuum Limit}

In the previous section starting from the DLS we have obtained,
near criticality, a continuous linear system
$$\eqalignno{
&(\partial^2+u){\tilde \xi}={\tilde \lambda}{\tilde \xi} &(7.1a)\cr
&{{\partial}\over{\partial{\tilde t_n}}}{\tilde \xi}=
(\partial^2+u)^{n+{1\over2}}_+
{\tilde \xi}&(7.1b)\cr
&{\tilde P}=\sum_{n=1}^{\infty}(n+{1\over2}){\tilde t_n}{\tilde Q}^{n+
{1\over2}}_+ &(7.4)\cr}
$$
whose consistency conditions are
$$\eqalignno{
[{\tilde Q}, {\tilde P}]&=1 &(7.2a)\cr
{{\partial}\over{\partial{\tilde t_n}}}u&=[(\partial^2+u)^{n+{1\over 2}}_+,
\partial^2+u] &(7.2b)\cr
{{\partial}\over{\partial{\tilde t_n}}}{\tilde P}
&=[(\partial^2+u)^{n+{1\over2}}_+, {\tilde P}]. &(7.2c)\cr}
$$

We want now to recover the Virasoro constraints in this continuous system.
The strategy is the same as for the discrete case. We use essentially the
string equation (7.2a). First of all, as is well known, in the
continuum limit the partition function behaves like
$$
\eqalign{\ln Z&=\sum_{k=1}^{N-1}(N-k)\ln R_k +{\rm constant\hbox{~~~} terms}
\cr
&\rightarrow\int^{\tilde t_0}_{0}d{\tilde t_0}^{'}
({\tilde t_0}-{\tilde t_0}^{'})u({\tilde t_0}^{'})+{\cal O}({\epsilon})
+{\rm regular\hbox{~~~}terms}\cr}
$$
$$
\Rightarrow \partial^2\ln {\tilde Z}=u({\tilde t_0}) \eqno (7.3)
$$
where, in taking the continuum limit, we passed to the
normalized partition function
$${\tilde Z}\equiv Z({\tilde t})/Z(T) \eqno (7.4)$$
the parameter $T$ being a reference point connected with the extremum of
integration $\tilde t_0=0$.

Now, using (7.1c),
eq.(7.2a) can be written
$$
-\sum_{k=1}^{\infty}(k+{1\over2}){\tilde t_k}
{\cal R}^{'}_k[u]=1.\eqno\hbox{(7.5)}
$$
Integrating once with respect to $t_0$, we obtain
$$
\sum_{k=1}^{\infty}(k+{1\over2}){\tilde t_k}\Big({\cal R}_k[u]-
{\cal R}_k[0]\Big)+{\tilde t_0}=0 \eqno (7.6)
$$
where ${\cal R}_k[0]$ is ${\cal R}_k[u]$ computed at
${\tilde t_0}=0$
\footnote{If one directly considers the continuum limit
of the discrete Virasoro constraints, it is not difficult to verify,
at least for the first few critical points,
that this is actually the case.}.
In order to simplify the next formulas let us introduce
the recursion operator
$$
{\hat\phi}\equiv {1\over4}\partial^2+u+{1\over2}u^{'}
\partial^{-1}
$$
and define the recursion relation for the Gelfand--Dickii
polynomials
$$
{\cal R}_{n+1}^{'}={\hat\phi}{\cal R}_n^{'}={\hat\phi}^n\partial u.
\eqno\hbox{(7.7)}
$$
Remembering that, on the basis of our conventions, we have
$$
\partial^{-1} {\cal R}_k'[u]= {\cal R}_k[u]-{\cal R}_k[0]
$$
eq.(7.6) can be rewritten
$$
F\equiv {\tilde t_0}+\sum_{k=1}^{\infty}(k+{1\over2}){\tilde t_k}
\partial^{-1}{\hat\phi}^{k-1}\partial u=0.\eqno\hbox{(7.8a)}
$$
On the same basis we can write
$$
\partial^{-1}{\hat\phi}^{n+1}\partial F=0, \qquad\qquad n\geq -1
\eqno\hbox{(7.8b)}
$$

To obtain these equations
we have used only the string equation. We can as well envisage eqs.(7.8)
as a consequence of a symmetry of the system, precisely as a consequence
of the fact that $u({\tilde t_0})$ and the KdV hierarchy
are invariant under the transformations
$$\cases{{\tilde t_k}\longrightarrow{\bar t_k}={\tilde t_k}+\epsilon
(k-n+{1\over2}){\tilde t_{k-n}}\cr\noalign{\vskip12pt}
u({\tilde t})\longrightarrow {\hat\phi}^{n+1}\cdot 1+u({\bar t_k})
\cr}.\eqno\hbox{(7.9)}
$$
The generators associated with (7.9) are
$$\eqalign{
L_{-1}&=
\sum_{k=1}^{\infty}(k+{1\over2}){\tilde t_k}
{{\partial}\over{\partial{\tilde t_{k-1}}}}
+{1\over {4\rho}}{\tilde t}^2_0,\cr\noalign{\vskip12pt}
L_0&=\sum_{k=0}^{\infty}(k+{1\over2}){\tilde t_k}
{{\partial}\over{\partial{\tilde t_k}}}+{1\over16},
\cr\noalign{\vskip12pt}
L_n&=\sum_{k=0}^{\infty}(k+{1\over2}){\tilde t_k}
{{\partial}\over{\partial{\tilde t_{k+n}}}}
+{{\rho}\over{4}}\sum_{k=1}^n
{{\partial^2}\over{\partial{\tilde t_{k-1}}\partial{\tilde t_{n-k}}}}
,\qquad\qquad n\ge 1.
\cr}\eqno\hbox{(7.10)}
$$
Here, for later purposes, we have introduced a constant $\rho$ (for example,
by rescaling all the $\tilde t$'s).

Even though we will not use it in the following, it is worth mentioning that
there is a larger symmetry of the system: the latter is also
invariant under the transformations
$$
{\tilde t_k}\longrightarrow{\tilde t_k}+\epsilon\eqno\hbox{(7.11a)}
$$
whose generators are given by
$$
V_k={{\partial}\over{\partial{\tilde t_k}}}, \qquad\qquad\qquad k\geq 0.
\eqno\hbox{(7.11b)}
$$
The generators $V_k$'s and $L_n$'s characterize the master symmetry of the
KdV hierarchy [9,11]. The corresponding algebra is
$$
\eqalignno{&[V_k, V_l]=0,\qquad\qquad\qquad\qquad k, l\geq 0
,&(7.12a)\cr\noalign{\vskip12pt}
&[V_k, L_n]=(k+{1\over2})V_{k+n}, \qquad\quad k\geq 0, k+n\geq 0,
&(7.12b)
\cr\noalign{\vskip12pt}
&[V_0, L_{-1}]={1\over{2\rho}}{\tilde t_0}&(7.12c)\cr\noalign{\vskip12pt}
&[L_n, L_m]=(n-m)L_{n+m},\qquad\quad n, m\geq -1.&(7.12d)\cr}
$$

But let us return to the derivation of the Virasoro constraint.
Representing now (7.8) in terms of the partition function,
we get for $n=-1,0$
$$\eqalign{
\partial\bigg(\sum_{n=1}^{\infty}(n+{1\over2}){\tilde t_n}
{{\partial}\over{\partial{\tilde t_{n-1}}}}\ln {\tilde Z}
+{1\over{2 \rho}}{\tilde t}^2_0\bigg)&=0\cr
\partial\bigg(\sum_{n=0}^{\infty}(n+{1\over2}){\tilde t_n}
{{\partial}\over{\partial{\tilde t}_n}}\ln {\tilde Z}\bigg)&=0\cr}
$$
or, in general,
$$
\partial\bigg({{l_n\sqrt{\tilde Z}}\over{\sqrt{\tilde Z}}}\bigg)=0,
\qquad\qquad\qquad n\ge -1. \eqno\hbox{(7.13)}
$$
where, by definition,
$$
l_0=L_0-{1\over16}, \qquad l_n=L_n,\qquad n\neq 0.
$$
The most general solution of (4.28) has the form
$$
l_n\sqrt{\tilde Z}=b_n\sqrt{\tilde Z}\eqno\hbox{(7.14)}
$$
where the $b_n$'s are ${\tilde t_0}$--independent
but  arbitrary  functions of the other parameters.
In order to determine them we remark that
they must be compatible with the algebra (7.12). In particular
they must satisfy
$$
[l_n, b_m]-[l_m, b_n]=(n-m)b_{n+m}+{1\over8}n\delta_{n+m,0}.\eqno\hbox{(7.15)}
$$
Moreover we remark that (7.12) is a graded algebra provided we define
the degree as follows
$$
{\rm deg}[{\tilde t_k}]\equiv -k,\qquad {\rm deg}[
{{\partial}\over{\partial{\tilde t_l}}}]\equiv l,\qquad {\rm deg}[\rho]\equiv
1
$$
Therefore, ${\rm deg}[L_n]=n$ and, from (7.14),
$$
{\rm deg}[b_n]=n, \qquad\qquad\qquad n\geq -1\eqno (7.16)
$$
The general form of $b_n$ will be a sum of monomials of the following
type
$$
{\cal M}_n(p, q_1,...,q_a)
={\rm const}~ \rho^p~ \tilde t_{n_1}^{q_1}... \tilde t_{n_a}^{q_a}
$$
where p is a real number and $q_1, q_2,...$ are nonnegative real numbers
(we exclude negative exponents as it is natural to require a smooth limit of
$b_n$ as any one of the couplings vanishes).
Next we remember that the parameter $\rho$ appeared on the scene because
of a rescaling $\tilde t_n \rightarrow \sqrt \rho \tilde t_n$. Therefore
if we perform the opposite rescaling the dependence of $b_n$ on $\rho$
must disappear.
This implies the condition $p= \textstyle {1\over 2} (q_1+...+q_a)$.
So the degree of the above monomial would be
$$
{\rm deg}[{\cal M}_n(p, q_1,...,q_a)]=\sum_{i=1}^a ({1\over 2}- n_i)q_i
\eqno (7.17)
$$
Comparing eq.(7.16) with (7.17), we see
all the  $b_n$'s are zero except perhaps for
$b_0$ which must be a constant, and $b_{-1}$, which
could depend on $\tilde t_{1}$ and $\rho$.
We now use the consistency conditon (7.15). For $n=1$ and $m=-1$
it tells us that
$$
{5\over 2} \tilde t_2 {{\partial b_{-1}} \over {{\partial \tilde t_1}}}
= 2 b_0 +{1\over 8}
$$
This allows us to conclude that
$$
b_0=-{1\over16}.
$$
and $b_{-1}$ does not depend on $\tilde t_1$.
Next we use again (7.15) for $n=0$ and $m=-1$ and conclude that
$$
b_{-1}=0
$$
Collecting the above results we obtain the Virasoro constraints
$$
L_n\sqrt{\tilde Z}=0,\qquad\qquad n\geq -1\eqno\hbox{(7.18)}
$$
with $L_n$ given by (7.10). Another proof of the same result is given in
Appendix.

Finally we recall that due to (7.2b) and (7.3), $\sqrt{\tilde Z}$ is a
$\tau$-function of the KdV hierarchy.

\vskip 1cm
\centerline{\bf Appendix}
\vskip .2cm

This Appendix is devoted to another proof of eq.(7.18).
We start from eq.(7.15) for $m=-1$ and $m=0$ and analyze the $\tilde t_0$
dependence. Since all the $b_n$ are $\tilde t_0$--independent we get,
in the first case ($m=-1$),
$$
{{\partial b_{-1}} \over {{\partial \tilde t_n}}} =0,\quad\quad n\geq 1
$$
and, in the second case ($m=0$),
$$
{{\partial b_{0}} \over {{\partial \tilde t_n}}} =0,\quad\quad n\geq 1
$$
In conclusion both $b_0$ and $b_{-1}$ are constant. Applying now again
eq.(7.15)
for $n=0$ and $m=-1$, we obtain
$$
b_{-1}=0
$$
i.e. eq.(7.18) is true for $n=-1$.
Since we have shown above that $\sqrt {\tilde Z}$ is a $\tau$-function of the
KdV hierarchy, we can now apply a theorem of ref.[11] which asserts that
if $L_{-1} \tau =0$, then $L_n \tau =0,$ $\forall n\geq -1$. Consequently
eq.(7.18) is proven.

\vglue 2.0cm
\centerline{\bf References}
\vglue 0.5cm

\item{[1]}
E. Brezin and V. Kazakov, Phys.Lett.236B (90) 144;

M. Douglas and S. Shenker,
Nucl.Phys.B335 (90) 635;

D. Gross and A. Migdal,
Phys.Rev.Lett.64 (90) 127;

T. Banks, M. Douglas,
N. Seiberg and S. Shenker,
Phys.Lett.238B (90) 279;

M. Douglas,
Phys.Lett. 238B (90) 176.

\item{[2]} V.A.Kazakov, in Proc. Carg\'ese workshop in 2d gravity, eds.
O.Alvarez, E.Marinari and P.Windey;

L.Alvarez-Gaum\'e, Helv.Phys.Acta 64 (1991) 361;

A.Bilal, CERN-TH.5867/90.

\item{[3]} E.J.Martinec, ``On the Origin of Integrability of Matrix
Models'', Chicago preprint, EFI--90--67.

\item{[4]} A. Gerasimov, A. Marshakov, A. Mironov,
A. Morozov, and A. Orlov, ``Matrix Models of Two--Dimensional
Gravity and Toda Theory'',
ITEP preprint (1990).

\item{[5]} R. Dijkgraaf, H. Verlinde, and E. Verlinde Nucl.Phys.
B348(1991)435.

\item{[6]} M. Fukuma, H. Kawai and R. Nakayama, ``Continuum
Schwinger--Dyson Equations and Universal Structures in Two--Dimensional
Quantum Gravity'', Tokyo preprint
KEK--TH--251 (1990).

\item{[7]}
E. Brezin, C. Itzykson, G. Parisi, and J.-B. Zuber,
Comm.Math.Phys. 59 (78) 35;

\item{[8]} E. Witten, ``Two--Dimensional Gravity and Intersection Theory
 on Moduli Space'', IASSNS--HEP--90--45 (1990).

\item{[9]} L.D.Fadeev and L.Takhtajan, Lect.Notes in Phys. Vol. 246
(Springer, Berlin, 1986), 66;

O.Babelon, Phys.Lett. B215 (1988) 523;

A.Volkov, Theor.Math.Phys. 74 (1988) 135.

\item{[10]} HoSeong La, ``Symmetries in Nonperturbative 2-d Quantum
Gravity, Pennsylvania preprint, UPR--0432 T (1990); W.Oevel,
``Master symmetries:
weak action angle structure for hamiltonian and non-hamiltonian systems"
Paderborn preprint (1987); J.Goeree, ``W-constraints in two dimensional quantum
gravity" Utrecht preprint THU-199/90.

\item{[11]} V.Kac and A.Schwarz, ``Geometric Interpretation of Partition
Function of 2D Gravity" preprint.

\bye